\documentclass{ws-procs9x6}

\newcommand {\nc} {\newcommand}
\nc {\beq} {\begin{eqnarray}}
\nc {\eeq} {\end{eqnarray}}

\begin{document}

\title{Tetraquark and Pentaquark Systems in Lattice QCD\footnote{\uppercase{T}alk presented 
by \uppercase{F}umiko \uppercase{O}kiharu at ``\uppercase{Q}uark \uppercase{N}uclear \uppercase{P}hysics 2005" (\uppercase{QNP}05).}
}

\author{Fumiko Okiharu}

\address{
Department of Physics, Nihon University, \\
1-8-14 Kanda Surugadai, Chiyoda, Tokyo 101-8308, Japan}

\author{Takumi Doi}

\address{
RIKEN-BNL Research Center, BNL, Upton, New York 11973, USA}

\author{Hiroko Ichie, Hideaki Iida, Noriyoshi Ishii, Makoto Oka, Hideo Suganuma}

\address{
Faculty of Science, Tokyo Institute of Technology, 
Tokyo 152-8551, Japan}

\author{Toru T. Takahashi}

\address{
YITP, Kyoto University, 
Kitashirakawa, Sakyo, 
Kyoto 606-8502, Japan}

\maketitle

\abstracts
{Motivated by the recent experimental discoveries of 
multi-quark candidates, e.g., the $\Theta^+(1540)$, 
we study multi-quark systems in lattice QCD. 
First, we perform accurate mass measurements of 
low-lying 5Q states with $J=1/2$ and $I=0$ 
in both positive- and negative-parity channels in anisotropic lattice QCD. 
The lowest positive-parity 5Q state is found to have a large mass of about 
2.24GeV after the chiral extrapolation. 
To single out the compact 5Q state from $NK$ scattering states, 
we develop a new method with the hybrid-boundary condition (HBC), 
and find no evidence of the compact 5Q state below 1.75GeV in the negative-parity channel. 
Second, we perform the first study of the multi-quark potential in lattice QCD 
to clarify the inter-quark interaction in multi-quark systems. 
The 5Q potential $V_{\rm 5Q}$ for the QQ-${\rm \bar{Q}}$-QQ system 
is found to be well described by the ``OGE Coulomb plus multi-Y Ansatz": 
the sum of the one-gluon-exchange (OGE) Coulomb term 
and the multi-Y-type linear term based on the flux-tube picture. 
The 4Q potential $V_{\rm 4Q}$ for the QQ-${\rm \bar{Q}\bar{Q}}$ system 
is also described by the OGE Coulomb plus multi-Y Ansatz, 
when QQ and $\rm \bar Q \bar Q$ are well separated.  
The 4Q system is described as a ``two-meson" state with disconnected flux tubes, 
when the nearest quark and antiquark pair is spatially close.
We observe a lattice-QCD evidence for the ``flip-flop'', i.e., 
the flux-tube recombination between the connected 4Q state and the ``two-meson'' state.
On the confinement mechanism,  
the lattice QCD results indicate the flux-tube-type linear confinement 
in multi-quark hadrons.
}

\section{Introduction} 
Recently, several new particles were experimentally discovered 
as the candidates of multi-quark hadrons. 
The $\Theta^+$(1540) was found in 2002 at SPring-8\cite{LEPS} as a pentaquark (5Q) baryon 
and was confirmed in other experimental groups (ITEP, JLab, ELSA).\cite{DIANA,CLAS,SAPHIR}
The $\Theta^+(1540)$ has the baryon number $B=1$ and the strangeness $S=+1$, and 
hence it is a manifestly exotic baryon and is considered to be a pentaquark ($uudd\bar{s}$) in the valence-quark picture. 
For other candidates of pentaquarks, 
the $\Xi^{--}$(1862) (ddss$\bar{u}$) was found at CERN,\cite{NA49}
and the $\Theta_c$(3099) ($uudd\bar{c}$) was reported as a charmed pentaquark at HERA.\cite{H1} 
Subsequently, the candidates of tetraquark (4Q) mesons were observed. 
The $X$(3872)\cite{Belle1,CDF2,D0,BABAR1} was found in the process of 
$B^+ \to K^++X(3872) \to K^++\pi^-\pi^+J/\psi$ at KEK.\cite{Belle1} 
The $X$(3872) is much heavier than the $J/\psi$, 
and its mass is close to the threshold of $D^0$ $(c\bar{u})$ and $\bar{D}^{0*}$ $(u\bar{c})$. 
Nevertheless, its decay width is very narrow as $\Gamma <2.3$MeV (90\% C.L.). 
These features indicate the $X$(3872) to be a tetraquark, e.g., a bound state of $D^0$ and $\bar{D}^{0*}$. 
Similarly, the $D_s$(2317)\cite{BABAR2,Belle2} is expected to be a tetraquark candidate. 
A common remarkable feature of 
all these exotic hadrons is their extremely narrow width, 
which seems a new mystery appearing in hadron physics. 
In any case, these discoveries of multi-quark hadrons are expected 
to reveal new aspects of hadron physics. 

In the theoretical side, 
the quark model is one of the most popular models to describe hadrons. In the quark model, 
mesons and baryons are usually described as $q\bar q$ and $3q$ composite particles, respectively. 
In more microscopic viewpoint, quantum chromodynamics (QCD) is the fundamental theory to describe the strong interaction. 
In terms of QCD, not only ordinary $q\bar q$ mesons and $3q$ baryons, 
but also exotic hadrons, such as multi-quark hadrons ($q\bar qq\bar q, qqqq\bar q$, $\cdots$), 
hybrid mesons ($q\bar qg$, $\cdots$), hybrid baryons ($qqqg$, $\cdots$) and glueballs ($gg, ggg$, $\cdots$) are 
expected to appear. 
We here aim to study these multi-quark hadrons directly based on QCD. 
Even at present, however, it is rather difficult to deal with the low-energy region analytically in QCD owing to its strong-coupling nature.
As an alternative way,  
the lattice QCD Monte Carlo simulation is established as the powerful method to treat 
non-perturbative nature of low-lying hadrons including exotic hadrons. 
In this paper, we perform the following two lattice QCD studies to clarify the properties of multi-quark systems. 

First, we investigate the mass and the parity of the 5Q system in lattice QCD. 
As for the parity assignment of the $\Theta(1540)$, 
it is not determined experimentally yet, and little agreement is achieved even in the theoretical side. 
On one hand, the positive-parity assignment is supported by the chiral soliton model\cite{DPP97} and the diquark model.\cite{JW03} 
On the other hand, the negative-parity assignment is supported by the nonrelativistic quark model,\cite{SR03} 
the QCD sum rule\cite{SDO04} and so on. 
These predictions have been done with model calculations, but these models were originally constructed only for ordinary hadrons. 
In fact, it is nontrivial that these models can describe the multi-quark system beyond the ordinary hadrons.
To get solid information for the multi-quark systems,  
we study their properties directly from QCD by the lattice QCD simulation,\cite{C7980,R97}  
which is the first-principle calculation and model independent. 

Second, we study the inter-quark interaction in multi-quark systems in lattice QCD. 
The inter-quark force is one of the most important elementary quantities in hadron physics. 
Nevertheless, for instance, no body knows the exact form of the confinement force in the multi-quark systems directly from QCD. 
In fact, some hypothetical forms of the inter-quark potential have been used in almost all quark model calculations so far.
Then, the lattice QCD study of the inter-quark interaction is quite desired for the study of the multi-quark systems. 
It presents the proper Hamiltonian in multi-quark systems 
and leads to a guideline to construct the QCD-based quark model. 
In this paper, 
to clarify the inter-quark force in the multi-quark system, 
we study the static multi-quark potential systematically in lattice QCD 
using the multi-quark Wilson loop. 
We investigate the three-quark (3Q) potential,\cite{TMNS01,TSNM02,TS0304}
which is responsible to baryon properties, 
and perform the first lattice-QCD study for the multi-quark potential,  
the tetraquark (4Q) and the pentaquark (5Q) 
potentials.\cite{OST05,STOI05,OST05p,SOTI05,OST04}

We show in Fig.1 our global strategy to understand the hadron properties from QCD.
One way is the direct lattice QCD calculations for 
the low-lying hadron masses and simple hadron matrix elements, 
although the wave function is unknown and 
the practically calculable quantities are severely limited.
The other way is to construct the quark model from QCD.
From the analysis of the inter-quark forces in lattice QCD, 
we extract the quark-model Hamiltonian.
Through the quark model calculation, one can obtain 
the quark wave-function of hadrons and more complicated properties of hadrons 
including properties of excited hadrons.

\begin{figure}[h]
\centering
\includegraphics[height=5cm]{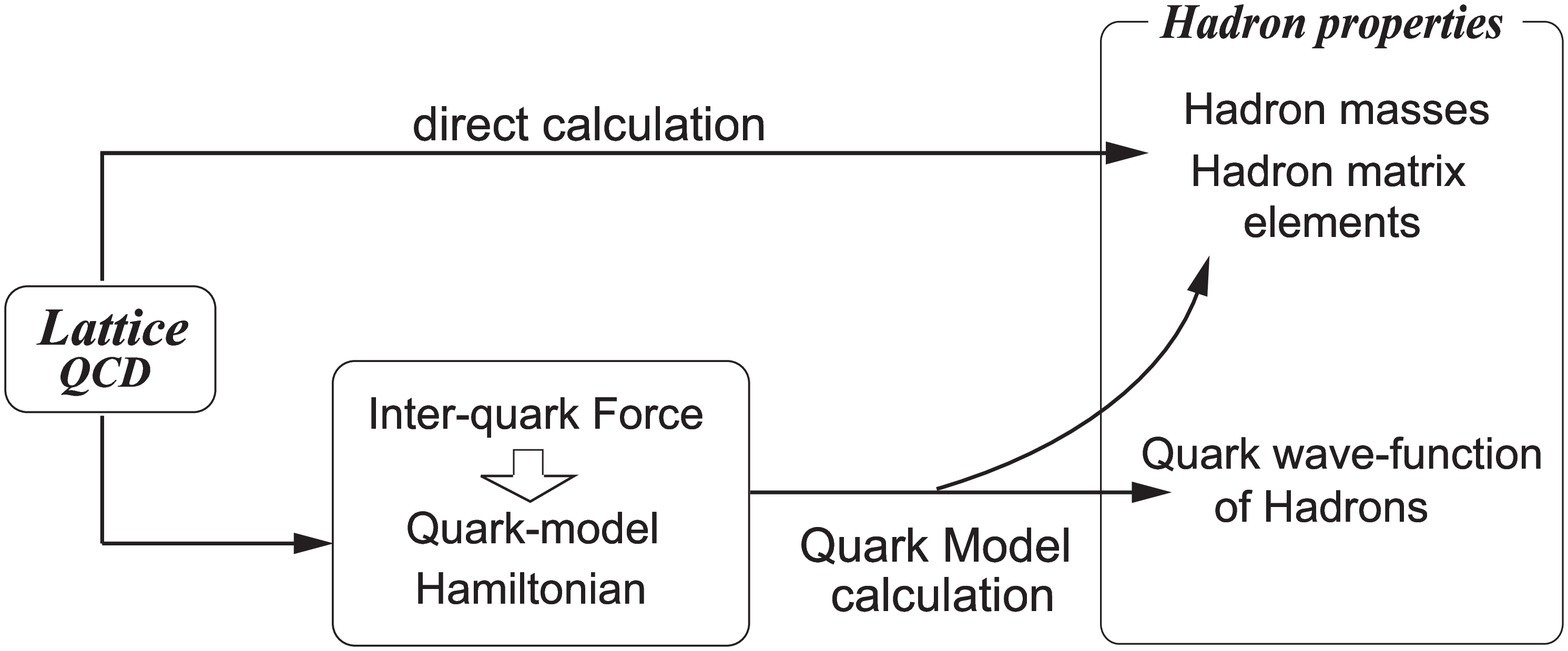}
\caption{Our global strategy to understand the hadron properties from QCD.
One way is the direct lattice QCD calculations for 
the low-lying hadron masses and simple hadron matrix elements, 
although the wave function is unknown and 
the practically calculable quantities are severely limited.
The other way is to construct the quark model from QCD.
From the analysis of the inter-quark forces in lattice QCD, 
we extract the quark-model Hamiltonian.
Through the quark model calculation, one can obtain 
the quark wave-function of hadrons and more complicated properties of hadrons 
including properties of excited hadrons.
}
\end{figure}

This paper is organized as follows. In Section 2, we perform accurate mass calculations 
for low-lying 5Q systems in anisotropic lattice QCD.\cite{IDIOOS05} 
In Section 3, we perform the systematic study of the inter-quark interaction in multi-quark 
systems.\cite{OST05,STOI05,OST05p,SOTI05,OST04} 
Section 4 is devoted for the summary and concluding remarks.

\section{Lattice QCD Study for the $\Theta^+$(1540)}

There are many theoretical studies of the $\Theta^+(1540)$
in order to clarify its physical properties.\cite{Z04,O04} 
For instance, as was briefly mentioned in the previous section, 
its parity is not yet determined experimentally, and is still unsettled also in the theoretical side.
Furthermore, the existence of the $\Theta^+$(1540) itself is not experimentally established as a pentaquark resonance. 
In fact, recent some high-energy experimental groups reported 
no evidence of the $\Theta^+$(1540),\cite{ALEPH,BES,HERA-B} 
while several low-energy experimental groups reported its existence. 

Also in lattice QCD, there is no consensus on the existence and the parity assignment of the $\Theta^+(1540)$.
Two early works supported the negative-parity state for the $\Theta^+(1540)$,\cite{CFKK03,S04}
while one early work supported the positive-parity state.\cite{CH04}
(To be strict, the Sasaki's result is incredible since he did not include the exchange diagrams.)
Two recent groups indicate no evidence for the low-lying pentaquark narrow resonance,\cite{IDIOOS05,MLAB04}
and one recent study suggests a negative-parity pentaquark state 
in more highly-excited region around 1.8GeV.\cite{TUOK05}
In this section, we perform the accurate mass measurement of the 5Q system 
in anisotropic lattice QCD, and apply a new technique to distinguish 
a compact resonance and a scattering state.

\subsection{Strategy for High Precession Measurements in Lattice QCD}

As a difficulty on the lattice study of the $\Theta^+$, 
several $NK$ scattering states appear apart from the compact 5Q resonance state. 
In this paper, we use the term of the $\Theta^+$ only for the compact 5Q resonance to distinguish it from the $NK$ scattering state.
In order to examine whether the low-lying 5Q state appears as a compact resonance $\Theta^+$, 
we perform the accurate lattice QCD calculations with adopting the following three advanced methods.\cite{IDIOOS05}

\subsubsection{Usage of Anisotropic Lattice QCD}
We use the anisotropic lattice, where the temporal lattice spacing $a_t$ 
is much finer than the spatial one $a_s$ as shown in Fig.2. 
In lattice QCD, hadron masses are calculated from the asymptotic temporal behavior of the hadron correlators.
On the anisotropic lattice, 
we can get the detailed information on the temporal behavior of  
the 5Q correlator, and hence we can perform accurate mass measurements for the low-lying 5Q system. 

\begin{figure}[h]
\centering
\includegraphics[width=0.4\textwidth]{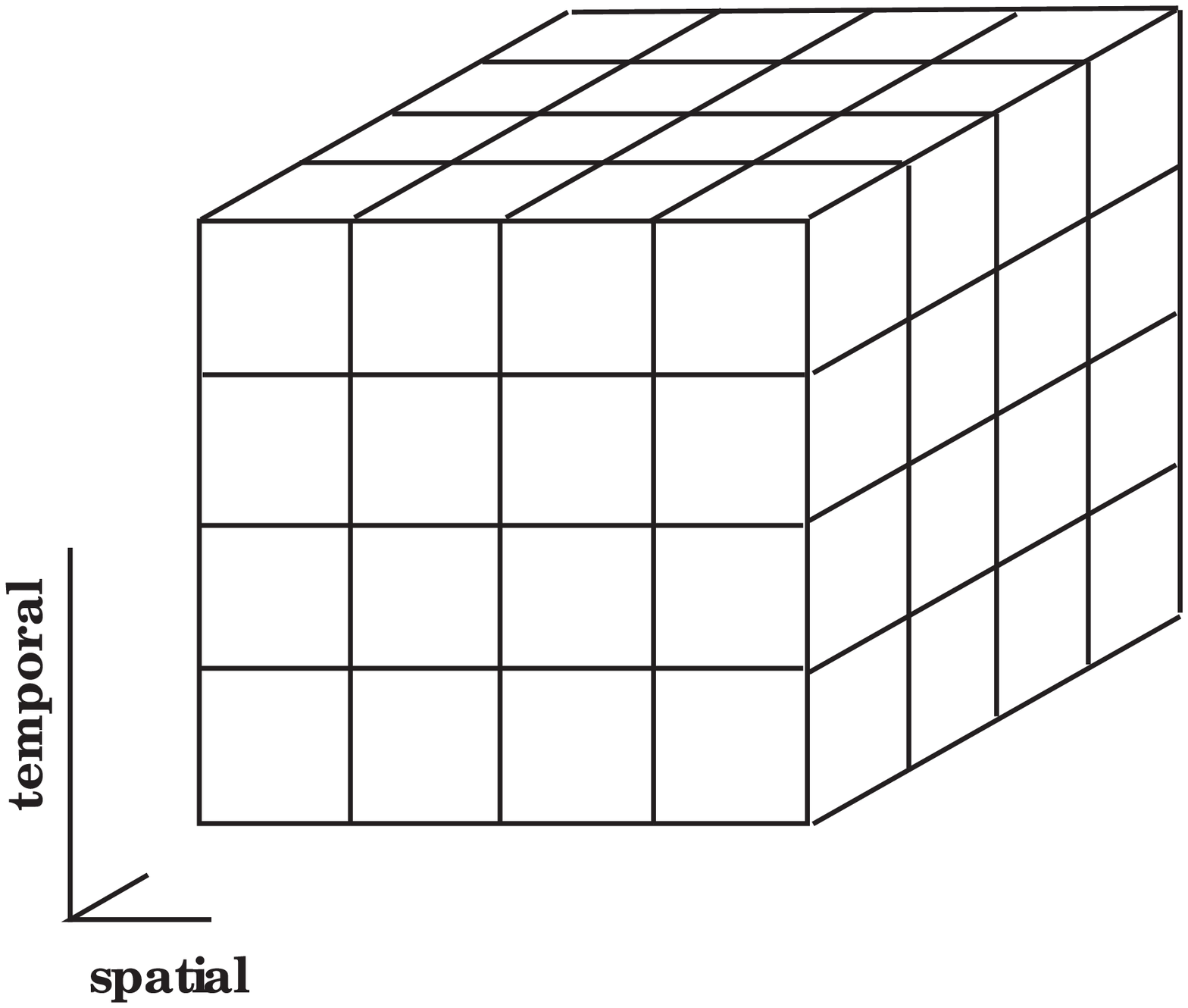} \hspace{0.5cm}
\includegraphics[width=0.4\textwidth]{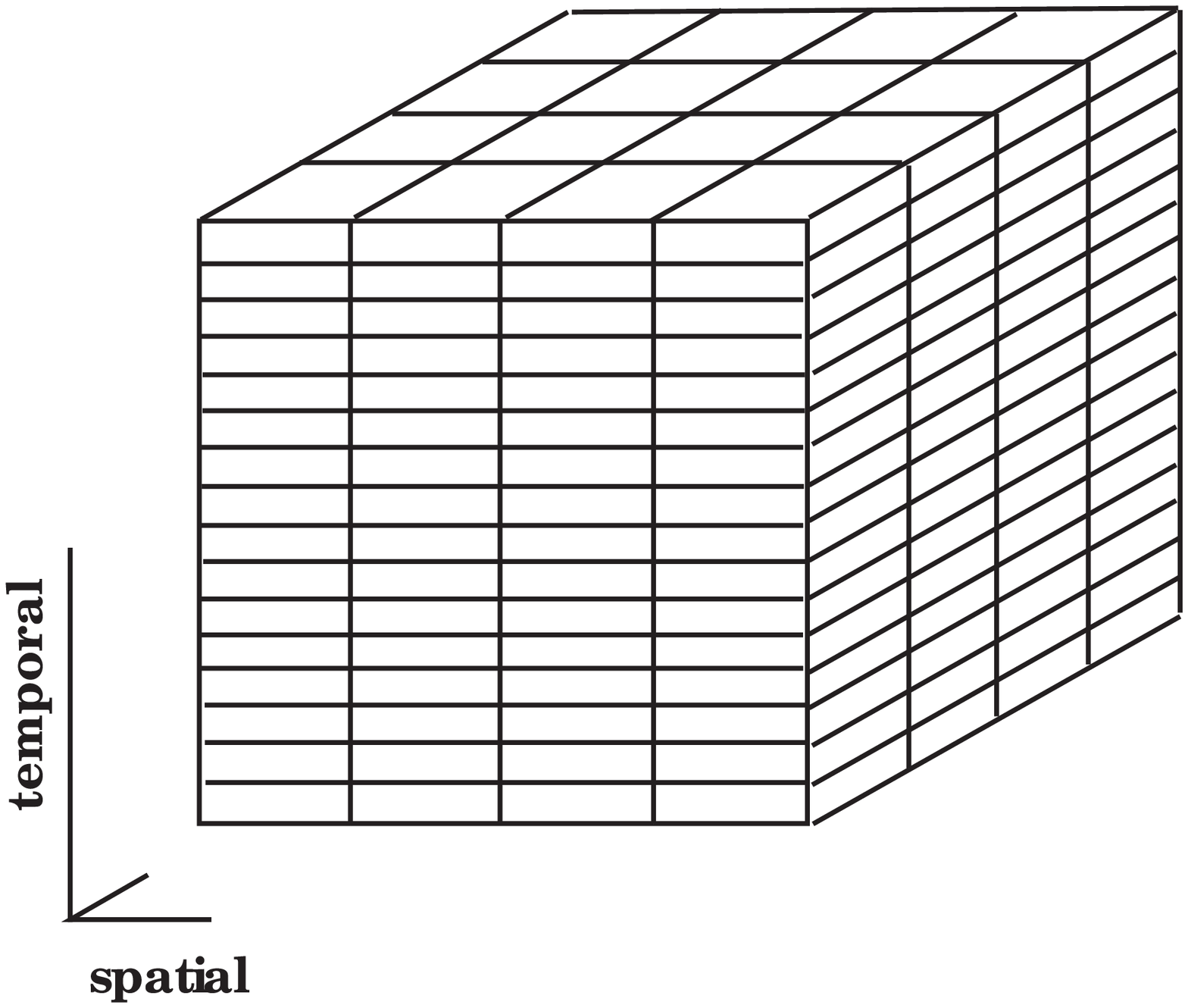}
\caption{Schematic figures of the isotropic lattice (left) and the anisotropic lattice (right).
On the anisotropic lattice, the temporal lattice spacing $a_t$ is taken to be smaller than the spatial one $a_s$.}
\end{figure}

\subsubsection{Usage of the Non-$NK$-type Interpolating Field Operator}
We use a non-$NK$-type interpolating field to extract the $\Theta^+$(1540) state. 
This choice of the interpolating field would be important and effective. 
In Ref.\refcite{MLAB04}, the authors used the $NK$-type interpolating field and only obtained 
the $NK$ scattering state instead of the compact 5Q state. 
However, their null result may be merely due to a small amount of the compact 5Q component 
in the $NK$-type interpolating field, because their calculation suffers from a large contamination of $NK$ scattering states.

We adopt the non-$NK$-type interpolating field,\cite{SDO04} 
\beq
O_{\alpha}\equiv 
\epsilon_{abc}\epsilon_{ade}\epsilon_{bfg}
(u^T_dC\gamma_5 d_e)(u^T_fCd_g)(C\bar{s}^T_c)_{\alpha}, 
\label{nonNK}
\eeq
for the 5Q state with spin $J=1/2$ and isospin $I=0$.
Here $\alpha$ denotes the Dirac index, and roman indices $a$-$g$ are color indices. 
$C \equiv \gamma_4\gamma_2$ denotes the charge conjugation matrix. 
Note that the non-$NK$-type operator in Eq.(\ref{nonNK}) 
cannot be decomposed into $N$ and $K$ in the nonrelativistic limit and its coupling to the $NK$ state is rather weak. 
Hence, the 5Q resonance state $\Theta^+$ can be singled out as much as possible 
in the present calculation, and the results are less biased by the contamination from $NK$ scattering states. 

\subsubsection{Application of the Hybrid Boundary Condition Method}
To distinguish compact resonances from scattering states, 
we propose a new method with the ``hybrid boundary condition" (HBC) instead of the ordinary 
periodic boundary condition.\cite{IDIOOS05} 
In the HBC, we impose the anti-periodic boundary condition for quarks ($u$, $d$) and 
the periodic boundary condition for antiquarks ($\bar s$), as shown in Table~1.
By applying the HBC on a finite-volume lattice, the $NK$ threshold is raised up, 
while the mass of a compact 5Q resonance $\Theta^+$ is almost unchanged.  
Therefore, we can distinguish a compact 5Q state $\Theta^+$ from an $NK$ scattering state 
by comparing between the HBC and the standard periodic boundary condition.

\begin{table}[htb]
\newcommand{\cc}[1]{\multicolumn{1}{c}{#1}}
\tbl{
The {\em hybrid  boundary  condition (HBC)} to distinguish a compact 5Q resonance $\Theta^+$
and an $NK$ scattering state for the $uudd\bar s$ system.
The standard boundary condition (BC) is also shown for comparison.
\vspace*{1pt}}
{\footnotesize
\begin{tabular}{llll} \hline \hline
  & u quark & d quark & s quark \\
\hline 
HBC & anti-periodic & anti-periodic & periodic \\
standard BC & periodic & periodic & periodic \\
\hline\hline
\end{tabular}
}
\end{table}

In lattice QCD with the finite spatial volume $L^3$, the spatial momenta are quantized as 
$p_i=2n_i\pi/L$ ($n_i \in {\bf Z}$) under the periodic boundary condition 
and $p_i=(2n_i+1)\pi/L$ under the anti-periodic boundary condition. 
In the periodic boundary condition, 
$N$ and $K$ can have zero momenta $|{\bf p}_{\rm min}|=0$ in the s-wave $NK$ scattering state.
The HBC imposes the anti-periodic boundary condition for $u$ and $d$ quarks and periodic boundary condition for $s$ quark, 
while the periodic boundary condition is usually employed for all $u$, $d$, $s$ quarks. 
In the HBC, the net boundary conditions of both $N$($uud,udd$) and $K$($u\bar{s},d\bar{s}$) are anti-periodic.
Then, under the HBC, $N$ and $K$ have minimum momenta $p_i=\sqrt{3}\pi/L$ in a finite box with $L^3$, 
and the threshold of the s-wave $NK$ scattering state is raised up 
as $\sqrt{m_N^2+{\bf p}_{\rm min}^2}+\sqrt{m_K^2+{\bf p}_{\rm min}^2}$. 
In contrast to $N$ and $K$, 
the compact 5Q resonance $\Theta^+$ ($uudd\bar s$) 
contains even number of $u$ and $d$ quarks, 
and hence its mass does not shift in the HBC. (See Table~2.)

\begin{table}[htb]
\newcommand{\cc}[1]{\multicolumn{1}{c}{#1}}
\tbl{
The net boundary condition for $\Theta^+$ ($uudd\bar s$), 
$N$ ($uud$ or $udd$) and $K$ ($d\bar s$ or $u\bar s$) 
in the hybrid boundary condition (HBC) and in the standard boundary condition (BC).
\vspace*{1pt}}
{\footnotesize
\begin{tabular}{llll} \hline \hline
  & $\Theta^+ (uudd\bar s)$ & $N (uud$ or $udd)$ & $K (d\bar s$ or $u\bar s)$ \\
\hline 
HBC & periodic & anti-periodic & anti-periodic \\
standard BC & periodic & periodic & periodic \\
\hline\hline
\end{tabular}
}
\end{table}

\subsection{Lattice QCD Setup for the Pentaquark Mass}

To generate gluon configurations, we use the standard plaquette action on the anisotropic lattice as~\cite{IDIOOS05} 
\beq
S_{\rm G}=\frac{\beta}{N_c}\frac1{\gamma_{\rm G}}  \sum_{s,i<j\le3}\mbox{Re} 
\mbox{Tr} \left\{ 1 - P_{ij}(s)\right\}+\frac{\beta}{N_c}\gamma_{\rm G}  
\sum_{s,i\le 3}\mbox{Re} \mbox{Tr}\left\{ 1 - P_{i4}(s)\right\},~~
\eeq
with $\beta \equiv 2N_c/g^2$, the plaquette $P_{\mu\nu}(s)$ and the bare anisotropy $\gamma_{\rm G}$. 

For the quark part, we adopt the $O(a)$-improved Wilson (clover) fermion action on the anisotropic lattice as 
\beq
S_{\rm F}&\equiv&\sum_{x,y} \bar\psi(x) K(x,y) \psi(y),
\eeq
with the quark kernel $K(x,y)$ as  
\beq
K(x,y)
&\equiv&
  \delta_{x,y}
  -\kappa_{t}\left\{
   (1 - \gamma_4)\; U_4(x)\; \delta_{x+\hat 4,y}
  +(1 + \gamma_4)\; U_4^\dagger(x - \hat 4)\; \delta_{x-\hat 4,y}
  \right\}
  \nonumber \\
&-&
  \kappa_{s}\sum_i
  \left\{(r - \gamma_i)\; U_i(x)\; \delta_{x+\hat i,y}
  +(r + \gamma_i)\; U_i^\dagger(x - \hat i)\; \delta_{x-\hat i,y}
  \right\} 
  \nonumber \\
&-&
  \kappa_{s}\; c_E \sum_i \sigma_{i4} G_{i4} \delta_{x,y}-
  r\; \kappa_{s}\; c_B \sum_{i<j} \sigma_{ij} G_{ij} \delta_{x,y},
\eeq
where $\kappa_{s}$ and $\kappa_{t}$ denote the spatial and temporal
hopping parameters, respectively. 
$G_{\mu\nu}$ denotes the field strength, 
which is defined through the standard clover-leaf-type construction.
The Wilson parameter $r$ and the clover coefficients, $c_{E}$ and $c_{B}$, 
are fixed by the tadpole-improved tree-level values as 
$r=1/\xi$, $c_E=1/(u_su_t^2)$ and $C_B=1/u_s^3$, where 
$u_s$ and $u_t$ denote the mean-field values of the spatial and the temporal link-variables, respectively.

For the lattice QCD simulation, 
we use $\beta=5.75$ and $12^3\times 96$ with the renormalized anisotropy $a_s/a_t=4$, 
which corresponds to $\gamma_{\rm G}=3.2552$. 
In this calculation, the lattice spacing is found to be  
$a_s\simeq$ 0.18fm $\simeq$ $(1.1{\rm GeV})^{-1}$ and $a_t\simeq$ 0.045fm $\simeq$ $(4.4{\rm GeV})^{-1}$. 
We adopt four values of the hopping parameter as $\kappa=0.1210~(0.0010)~0.1240$ for $u$ and $d$ quarks, 
and use $\kappa_{s\hbox{-}{\rm quark}}=0.1240$ for the $s$ quark.
We calculate typical hadron masses at each $\kappa$ as shown in Table~3, and 
find $\kappa_{\rm phys.}\simeq 0.1261$ corresponding to the physical situation of $m_{\pi} \simeq 0.14$ GeV.

\begin{table}[htb]
\newcommand{\cc}[1]{\multicolumn{1}{c}{#1}}
\tbl{
Masses of $\pi$, $\rho$, $K$ and $N$ 
for each hopping parameter $\kappa$ in the physical unit of GeV.  
$\kappa_{\rm phys.}\simeq 0.1261$ corresponds to the physical situation of $m_{\pi} \simeq 0.14$ GeV.
\vspace*{1pt}}
{\footnotesize
\begin{tabular}{llllll} \hline \hline
$\kappa$   & 0.1210 & 0.1220 & 0.1230 & 0.1240 & $\kappa_{\rm phys.}$ 
\\ 
\hline
$m_{\pi}$  & 1.005(2) & 0.898(2) & 0.784(2) & 0.656(3) & 0.140    \\
$m_{\rho}$ & 1.240(3) & 1.161(3) & 1.085(4) & 1.011(5) & 0.850(7) \\
$m_{K}$    & 0.845(2) & 0.785(2) & 0.723(2) & 0.656(3) & 0.530(4) \\
$m_{N}$    & 1.878(5) & 1.744(5) & 1.604(5) & 1.460(6) & 1.173(9) \\
\hline\hline
\end{tabular}
}
\end{table}

\subsection{Lattice QCD Results for the $\Theta^+$(1540)}

Now,  using anisotropic lattice QCD, 
we perform the accurate mass measurement of the low-lying 5Q states with $J=1/2$ and $I=0$ 
in both positive- and negative-parity channels 
from the correlator of the non-$NK$-type 5Q operator with parity projection.\cite{IDIOOS05} 

In Fig.3, we show the lattice QCD results for the masses of 
lowest positive- and negative-parity 5Q states under the standard periodic boundary condition. 
After the chiral extrapolation, 
the lowest positive-parity 5Q state is found to be rather heavy 
as $m_{\rm 5Q}(J^\pi=\frac12^+) \simeq$ 2.24 GeV, which 
seems to be too heavy to be identified as the $\Theta^+(1540)$.

\begin{figure}[htb]
\centering
\includegraphics[angle=-90,width=0.75\textwidth]{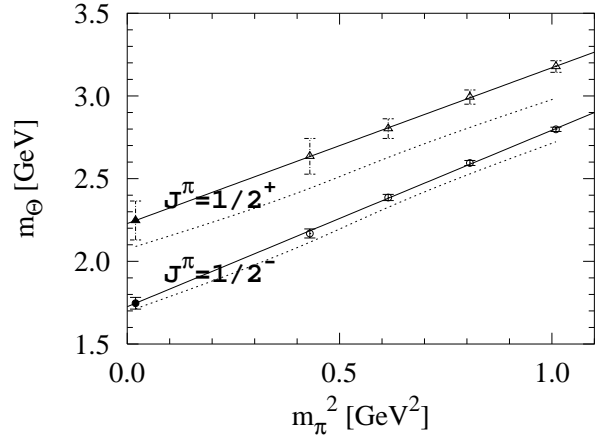}
\caption{The lowest mass $m_{\rm 5Q}$ of the positive- and negative-parity 5Q states plotted  against $m_{\pi}^2$. 
The open symbols denote the direct lattice QCD data for positive-parity (triangles) and negative-parity (circles).
The solid symbols denote the results of the chiral extrapolation. 
The dotted lines indicate the $NK$ thresholds for p-wave (upper) 
and s-wave (lower) cases.}
\end{figure}

On the other hand, we get a lower mass for the negative-parity 5Q state as 
$m_{\rm 5Q} (J^\pi=\frac12^-) \simeq$ 1.75 GeV after the chiral extrapolation. 
This value $m_{\rm 5Q} \simeq$ 1.75 GeV seems to be closer to 
the experimental result of $m_{\Theta^+} \simeq $ 1.54 GeV.
At this stage, however, this lowest negative-parity 5Q state may be merely an $NK$ scattering state, 
although the non-$NK$-type 5Q operator used in this calculation 
includes only a small amount of the $NK$ component.

To clarify whether the observed low-lying 5Q state is a compact 5Q resonance $\Theta^+$ or an $NK$ scattering state, 
we apply the new method with the hybrid-boundary condition (HBC), and 
compare the lattice results with the HBC and those with  
the standard periodic boundary condition (BC). 
Recall that, in the HBC, the $NK$ threshold is largely raised up, 
while the mass of the compact 5Q resonance ($uudd\bar s$) is to be almost unchanged, 
as was mentioned in Sect. 2.1.3. 

\begin{figure}[ht]
\centering
\includegraphics[angle=-90,width=0.65\textwidth]{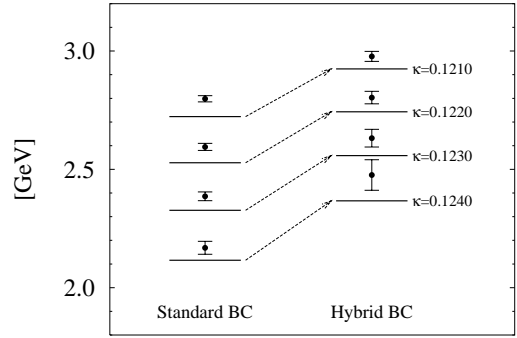}
\caption{
Comparison between the standard periodic boundary condition (Standard BC) and 
the Hybrid Boundary Condition (HBC) for the lowest mass of the negative-parity 5Q system.
At each $\kappa$, the lattice QCD result (the solid symbol) is raised up in accordance with 
the corresponding $NK$ threshold (the solid line).
This behavior indicates that the low-lying negative-parity 5Q state observed in lattice QCD 
is an NK scattering state rather than a compact 5Q resonance $\Theta^+$.
}
\end{figure}

In Fig.4, we show the mass of the lowest-lying negative-parity 5Q state in lattice QCD 
with the standard periodic BC and the HBC at each $\kappa$. 
The symbols denote the lattice QCD results for the 5Q state 
and the lines denote the $NK$ threshold at each $\kappa$. 
The left and right figures show the results with the standard periodic BC  
and the hybrid boundary condition (HBC), respectively. 
Note that the $NK$ threshold is estimated to be raised up about 200$-$250 MeV in the HBC.
  
As a lattice QCD result, the mass of the 5Q state is largely raised in the HBC in accordance with the $NK$ threshold, 
which indicates that the lowest 5Q state observed on the lattice is merely an s-wave $NK$ scattering state.
In other words, if there exists a compact 5Q resonance $\Theta^+$ below 1.75GeV, 
it should be observed in this lattice calculation with the non-$NK$-type operator, 
and its mass should be almost unchanged also in the HBC.  
However, there is no such a 5Q state observed in the lattice calculation, 
which means absence of the compact 5Q resonance $\Theta^+$ below 1.75GeV. 

To conclude, the present lattice QCD calculation at the quenched level seems to 
indicate absence of the low-lying compact 5Q resonance 
$\Theta^+$ with $J=1/2$ and $I=0$ near 1.54GeV.\cite{IDIOOS05} 

\subsection{Discussion on Null Result of $\Theta(1540)$ in Lattice QCD}

In this subsection, we consider the physical consequence of the present null result 
on the low-lying 5Q resonance $\Theta^+$ in lattice QCD.
One possible answer is absence of the pentaquark resonance $\Theta^+(1540)$, as is indicated by recent some high-energy experiments. 
However, if the $\Theta^+(1540)$ really exists, we deduce the following possibilities. 

\subsubsection{Quenching Effects}
First, the present lattice simulation has been done at the quenched level, where dynamical quark effects are suppressed. 
This quenching effect is not clear and then it may cause the 5Q resonance $\Theta^+$ to be heavier as an unknown effect.
Actually, recent one lattice study indicates a negative-parity 5Q resonance $\Theta^+$ 
in a higher-excited region of about 1.8GeV.\cite{TUOK05}

\subsubsection{Nontrivial Quantum Numbers of the $\Theta(1540)$}
Second, we have considered only the 5Q state with spin $J=1/2$ and isoscalar $I=0$.
However, the $\Theta^+(1540)$ may have other quantum numbers,\cite{CPR03,H03,KMN05}
 e.g., spin $J=3/2$, isovector $I=1$ or isotensor $I=2$.
Considering such a possibility, our group started to investigate the 5Q system with higher spin $J=3/2$ in lattice QCD as a next step.

\subsubsection{More Complicated Structure of the $\Theta^+(1540)$}
Third, we have used a localized 5Q interpolating field in this lattice QCD calculation. 
However, the actual $\Theta^+(1540)$ may have more complicated structure beyond the localized interpolating field. 
Such a possibility has been pointed out in the theoretical side.
For instance, Karliner and Lipkin\cite{KL03} proposed the diquark-triquark ($qq$-$qq\bar q$) picture for the $\Theta$, and 
Bicudo~et~al.\cite{BM04} pointed out the possibility of the heptaquark picture, where the $\Theta^+$ is 
described as a bound state of $\pi$, $K$ and $N$.
If the $\Theta^+(1540)$ has such a complicated structure, we have to use the corresponding nonlocal interpolating field to get its proper information.

\subsection{Necessity of the Wave Function of Pentaquarks}

So far, we have performed the direct mass measurement of 5Q states in lattice QCD, 
where the path integral over arbitrary states is numerically calculated on a supercomputer. 
In the path-integral formalism, however, 
it is rather difficult to extract the state information, such as the wave-function of the 5Q state, 
and therefore only limited simple information can be obtained in the direct lattice-QCD calculation.

Actually, to distinguish the compact 5Q resonance $\Theta^+$ from $NK$ scattering states
was very difficult in lattice QCD, 
and hence we had to develop a new method with the hybrid boundary condition (HBC).
In this respect, if the wave function is obtained, one easily finds out whether it is a compact resonance state $\Theta^+$ or not.

Indeed, to get the wave function is very important to clarify the further various properties of the 5Q state such as 
the underlying structure and the decay width, which cannot be obtained practically only with the direct lattice-QCD calculation.

Then, apart from the direct lattice-QCD calculation, we have to seek the way to obtain the proper wave function of the 5Q state.
To do so, we need a proper Hamiltonian for the 5Q system based on QCD. 
One possible way in this direction is to construct the quark model from QCD, as was mentioned in Section 1. 
In the next section, we study the inter-quark interaction in multi-quark systems directly from QCD,  
and aim to construct the QCD-based quark-model Hamiltonian.

\section{Lattice QCD Study for the Inter-Quark Interaction in Multi-Quark Systems}

In this section, we perform the first study of 
the inter-quark interaction in multi-quark systems in 
lattice QCD,\cite{OST05,STOI05,OST05p,SOTI05,OST04} 
and seek for the QCD-based quark-model Hamiltonian to describe multi-quark hadrons.
The quark-model Hamiltonian consists of the kinetic term and the potential term,
which is not known form QCD in multi-quark systems.

As for the potential at short distances, 
the perturbative one-gluon-exchange (OGE) potential would be appropriate, 
due to the asymptotic nature of QCD. 
For the long-range part, however, there appears the confinement potential as a typical 
non-perturbative property of QCD, and its form is highly nontrivial 
in the multi-quark system. 

In fact, 
to clarify the confinement force in multi-quark systems 
is one of the essential points for the 
construction of the QCD-based quark-model Hamiltonian.
Then, in this paper, we investigate the multi-quark potential in lattice QCD, 
with paying attention to the confinement force in multi-quark hadrons. 

\subsection{The Three-Quark Potential in Lattice QCD}

So far, only for the simplest case of static Q$\bar{\rm Q}$ systems,
detailed lattice QCD studies have been done,  
and the Q$\bar{\rm Q}$ potential $V_{\rm Q\bar Q}$ is known to be 
well described by the Coulomb plus linear potential as~\cite{C7980,R97,TMNS01,TSNM02}
\beq
V_{\rm Q\bar Q}(r)=-\frac{A_{\rm Q\bar Q}}{r}+\sigma_{\rm Q\bar Q} r+C_{\rm Q\bar Q}
\label{VQQ}
\eeq
with $r$ being the inter-quark distance.

To begin with, we study three-quark (3Q) systems in lattice QCD
to understand the structure of baryons at the quark-gluon level. 
Similar to the derivation of the Q$\rm\bar{Q}$ potential from the Wilson loop, 
we calculate the 3Q potential $V_{\rm 3Q}$ from the 3Q Wilson loop 
$W_{\rm 3Q}$ in SU(3) lattice QCD with  
($\beta$=5.7, $12^3\times 24$),
($\beta$=5.8, $16^3\times 32$), 
($\beta$=6.0, $16^3\times 32$) and 
($\beta=6.2$, $24^4$) at the quenched level.
For more than 300 different patterns of spatially-fixed 3Q systems, 
we perform accurate and detailed calculations for the 3Q potential,\cite{TMNS01,TSNM02,TS0304,STOI05,SOTI05}  
and find that the lattice QCD data of the 3Q potential $V_{\rm 3Q}$
are well described by the Coulomb plus Y-type linear potential, i.e., Y-Ansatz,  
\beq
V_{\rm 3Q}=-A_{\rm 3Q}\sum_{i<j}\frac1{|{\bf r}_i-{\bf r}_j|}
+\sigma_{\rm 3Q} L_{\rm min}^{\rm 3Q}+C_{\rm 3Q},
\label{V3Q}
\eeq
within 1\%-level deviation.\cite{TMNS01,TSNM02,TS0304,STOI05,SOTI05}
Here,  $L_{\rm min}^{\rm 3Q}$ is the minimal total length of the color flux tube, 
which is Y-shaped for the 3Q system.

To demonstrate the validity of the Y-Ansatz, 
we show in Fig.5 the lattice QCD data of 
the 3Q confinement potential $V_{\rm 3Q}^{\rm conf}$, 
i.e., 3Q potential subtracted by the Coulomb part, 
plotted against Y-shaped flux-tube length $L_{\rm min}^{\rm 3Q}$.
For each $\beta$, clear linear correspondence is found between 3Q confinement potential 
$V_{\rm 3Q}^{\rm conf}$ and $L_{\rm min}^{\rm 3Q}$,
which indicates the Y-Ansatz for the 3Q potential.\cite{STOI05,SOTI05}

\begin{figure}[h]
\begin{center}
\includegraphics[height=6.7cm]{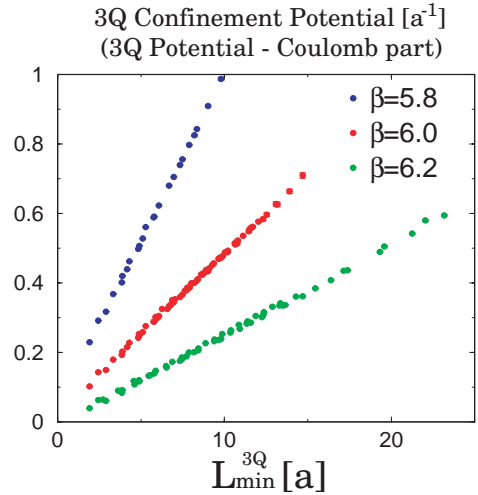} 
\caption{
The lattice QCD result for the 3Q confinement potential $V_{\rm 3Q}^{\rm conf}$, 
i.e., the 3Q potential subtracted by its Coulomb part, 
plotted against 
Y-shaped flux-tube length $L_{\rm min}^{\rm 3Q}$ 
at $\beta$=5.8, 6.0 and 6.2 in the lattice unit.
The clear linear correspondence between 3Q confinement potential 
$V_{\rm 3Q}^{\rm conf}$ and $L_{\rm min}^{\rm 3Q}$
indicates the Y-Ansatz for the 3Q potential.
}
\end{center}
\end{figure}

Here, we consider the physical meaning of the Y-Ansatz.
Apart from an irrelevant constant, 
the Y-Ansatz, Eq.(\ref{V3Q}), consists of the Coulomb term and the Y-type linear potential,
which play the dominant role at short and long distances, respectively.
The Coulomb term would originate from the one-gluon-exchange (OGE) process.
In fact, at short distances, perturbative QCD is applicable,
and therefore the inter-quark potential is expressed as 
the sum of the two-body one-gluon-exchange (OGE) Coulomb potential. 

The appearance of the Y-type linear potential supports 
the flux-tube picture\cite{N74,KS75,CNN79,CKP83} 
at long distances, where there appears  
the color flux tube linking quarks inside hadrons with its length minimized. 
In particular, the confinement force in baryons corresponds to 
the Y-shaped flux tube, which implies existence of the three-body interaction in baryons. 

In usual many-body systems, the main interaction is described by a two-body interaction 
and the three-body interaction is a higher-order contribution. 
In contrast, as is clarified by our lattice-QCD study, 
the quark confinement force in baryons is a genuinely three-body interaction,\cite{TMNS01,TSNM02} 
which is one of significant features of QCD. 
In fact, the appearance of the Y-type junction and the three-body confinement force 
reflect the SU(3) group structure in QCD, 
e.g., the number of color, $N_c=3$, and is peculiar to QCD.\cite{TMNS01,TSNM02}
In this sense, the study of the 3Q system is very important to get 
a deeper insight of the QCD physics.

\begin{figure}[ht]
\begin{center}
\includegraphics[height=4.6cm]{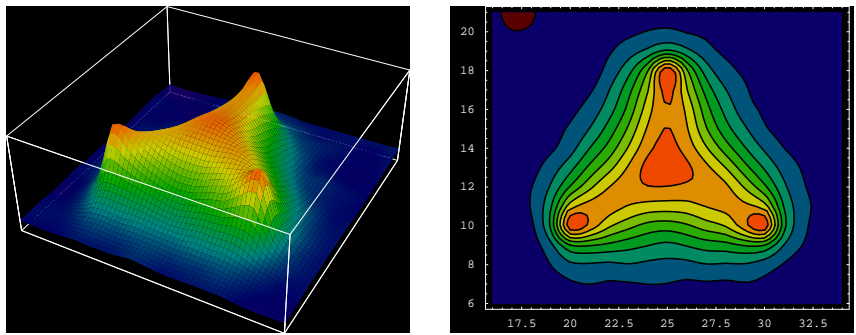} 
\caption{
Lattice QCD results for Y-shaped flux-tube formation in the 3Q system: 
a bird's-eye view and a contour map of 
the action density of the spatially-fixed 3Q system in maximally-Abelian projected QCD.
The distance between the junction and each quark is about 0.5 fm.
}
\end{center}
\vspace{-0.2cm}
\end{figure}

Recently, a clear Y-type flux-tube formation is actually observed 
for spatially-fixed 3Q systems in lattice QCD as shown in Fig.6.\cite{STOI05,SOTI05,IBSS03}
Thus, together with recent several other analytical and numerical 
studies,\cite{KS03,BS04,C0405}
the Y-Ansatz for the static 3Q potential is almost settled. 
This result indicates the color-flux-tube picture for baryons.

\subsection{The OGE Coulomb plus Multi-Y Ansatz}

Now, we proceed to multi-quark systems. 
We first consider the theoretical form of the multi-quark potential, 
since we will have to analyze the lattice QCD data by comparing them 
with some theoretical Ansatz.

By generalizing the lattice QCD result of the Y-Ansatz for the three-quark potential, 
we propose the one-gluon-exchange (OGE) Coulomb plus 
multi-Y Ansatz,\cite{OST05,STOI05,OST05p,SOTI05,OST04}
\beq
V=\frac{g^2}{4\pi}\sum_{i<j}\frac{T^a_i T^a_j}{|{\bf r}_i-{\bf r}_j|}+\sigma L_{\rm min}+C,
\label{VnQ}
\eeq
for the potential form of the multi-quark system.
Here, the confinement potential is proportional to the minimal total length $L_{\rm min}$ 
of the color flux tube linking the quarks, which is multi-Y shaped in most cases.

In the following, we study the inter-quark interaction in multi-quark systems 
in lattice QCD, and compare the lattice QCD data with the theoretical form in Eq.(\ref{VnQ}).
Note here that the lattice QCD data are meaningful as primary data 
on the multi-quark system directly based on QCD, and do not depend on any theoretical Ansatz.

\subsection{Formalism of the Multi-Quark Wilson Loop}

Next, we formulate the multi-quark Wilson loop to obtain 
the multi-quark potential in lattice QCD.\cite{OST05,STOI05,OST05p,SOTI05,OST04}

Similar to the derivation of the Q$\rm\bar{Q}$ potential from the Wilson loop, 
the static multi-quark potential can be derived from 
the corresponding multi-quark Wilson loop.  
We construct the tetraquark Wilson loop $W_{\rm 4Q}$ and the pentaquark Wilson loop $W_{\rm 5Q}$ in a gauge invariant manner 
as shown in Figs.7(a) and (b), respectively. 

\begin{figure}[ht]
\centering
\includegraphics[scale=0.33,clip]{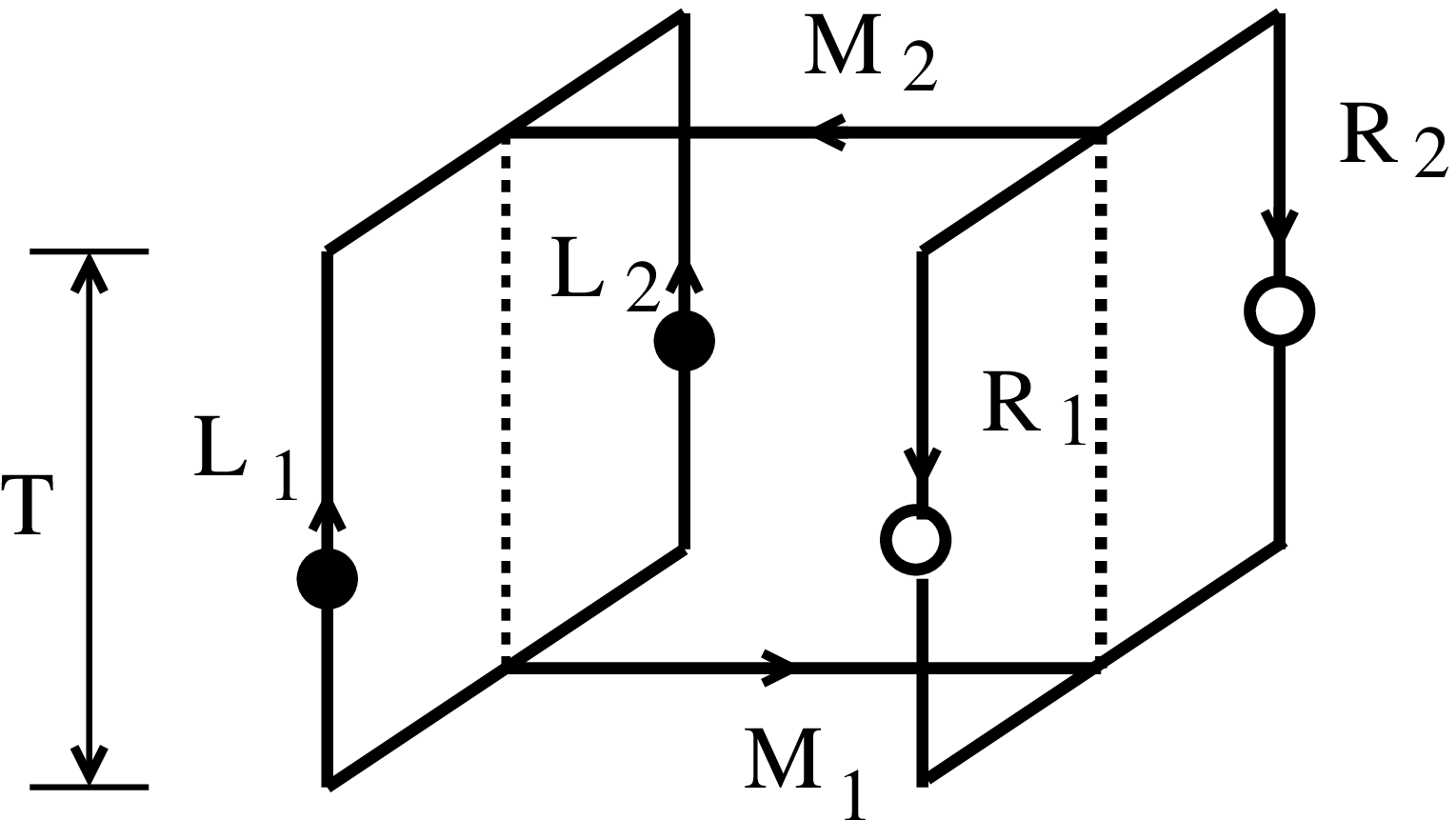}
\includegraphics[scale=0.33,clip]{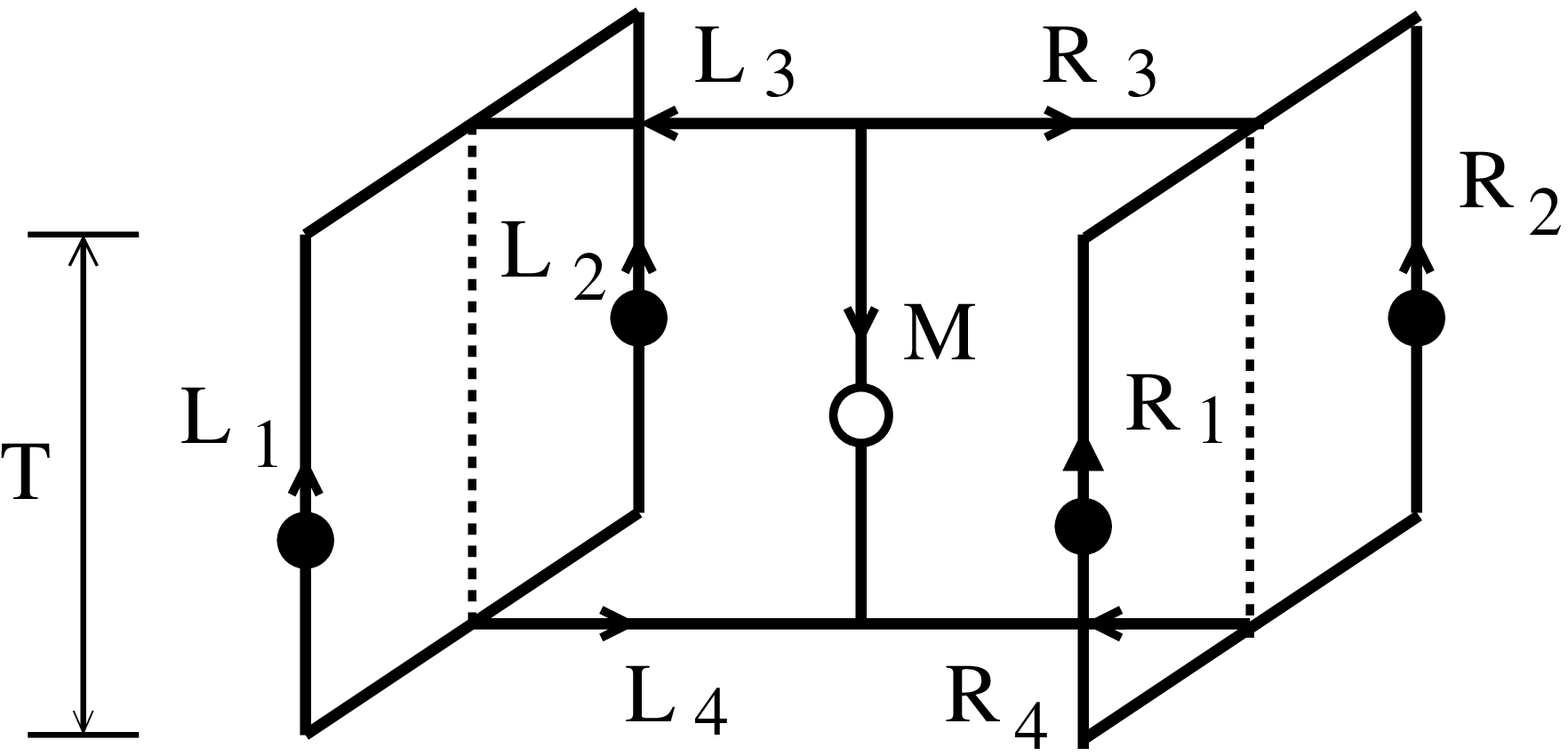}
\caption{(a) The tetraquark Wilson loop $W_{\rm 4Q}$. (b) The pentaquark Wilson loop $W_{\rm 5Q}$. The contours $M,M_i,R_j,L_j (i=1,2,j=3,4)$ are line-like and $R_j,L_j (j=1,2)$ are 
staple-like. The multi-quark Wilson loop physically means that 
a gauge-invariant multi-quark state is generated at $t=0$ and annihilated at $t=T$ with quarks being spatially fixed in ${\bf R}^3$ 
for $0<t<T$.}
\end{figure}

The tetraquark Wilson loop $W_{\rm 4Q}$ and the pentaquark Wilson loop $W_{\rm 5Q}$ are defined by 
\beq
W_{\rm 4Q}&\equiv& \frac{1}{3}{\rm tr}(\tilde{M}_1 \tilde{R}_{12} \tilde{M}_2 \tilde{L}_{12}),\nonumber \\
W_{\rm 5Q}&\equiv& \frac1{3!}\epsilon^{abc}\epsilon^{a'b'c'}M^{aa'}(\tilde R_3\tilde R_{12}\tilde R_4)^{bb'}(\tilde L_3\tilde L_{12}\tilde L_4)^{cc'}, 
\eeq
where $\tilde{M}$, $\tilde{M}_i$, $\tilde{L_j}$ and $\tilde{R_j}$ ($i$=1,2, $j$=1,2,3,4) are given by 
\beq
\tilde{M}, \tilde{M}_i, \tilde{R_j}, \tilde{L_j}
\equiv 
P \exp{\{ig \int_{M, M_i,R_j,L_j}dx^\mu A_\mu (x)\}}\in \rm{SU(3)_c}.
\eeq
Here, $\tilde{R}_{12}$ and $\tilde{L}_{12}$ are defined by
\beq
\tilde{R}_{12}^{a'a} 
\equiv \frac{1}{2}\epsilon^{abc}\epsilon^{a'b'c'}
R_1^{bb'}R_2^{cc'},\quad 
\tilde{L}_{12}^{a'a} 
\equiv \frac{1}{2}\epsilon^{abc}\epsilon^{a'b'c'}
L_1^{bb'}L_2^{cc'}.
\eeq
The multi-quark Wilson loop physically means that 
a gauge-invariant multi-quark state is generated at $t=0$ and annihilated at $t=T$ 
with quarks being spatially fixed in ${\bf R}^3$ for $0<t<T$.

The multi-quark potential is obtained from the vacuum expectation value of 
the multi-quark Wilson loop as
\beq
V_{\rm 4Q}=-\lim_{T\rightarrow \infty} \frac1{T} {\rm ln} \langle W_{\rm 4Q}\rangle, 
\qquad
V_{\rm 5Q}=-\lim_{T\rightarrow \infty} \frac1{T} {\rm ln} \langle W_{\rm 5Q}\rangle.
\eeq

\subsection{Lattice QCD Setup for the Multi-quark Potential}

Here, we briefly summarize the lattice QCD setup in this calculation.
For the study of the multi-quark potential,
the SU(3) lattice QCD simulation is done 
with the standard plaquette action at $\beta=6.0$ 
on the $16^3 \times 32$ lattice at the quenched level. 
(The calculation for large-size multi-quark configurations are performed 
by identifying $16^2 \times 32$ as the spatial size.)

In this calculation, the lattice spacing $a$ is estimated as $a \simeq$ 0.104fm, 
which leads to the string tension  $\sigma_{\rm Q\bar Q} = (427{\rm MeV})^2$ 
in the Q$\rm \bar Q$ potential.\cite{STOI05} 
We use 150 gauge configurations for the 5Q potential simulation and 300 gauge configurations for the 4Q potential simulation. 
The smearing method is used for the enhancement of the ground-state component. 
We here adopt $\alpha=2.3$ and the iteration number $N_{\rm smr}=40$, which lead to a large enhancement of 
the ground-state component.\cite{OST05,STOI05,OST05p,SOTI05,OST04}

\subsection{Lattice QCD Result of the Pentaquark Potential}

We perform the first study of the pentaquark potential $V_{\rm 5Q}$ in lattice QCD 
for 56 different patterns of QQ-$\rm \bar Q$-QQ type pentaquark configurations, 
as shown in Fig.8.
As the conclusion, the lattice QCD data of  
$V_{\rm 5Q}$ are found to be well described by the OGE Coulomb plus multi-Y Ansatz, i.e., 
the sum of the OGE Coulomb term and the multi-Y-type 
linear term based on the flux-tube picture.\cite{OST05,STOI05,OST05p,SOTI05} 

\begin{figure}[h]
\vspace{-0.5cm}
\begin{center}
\includegraphics[height=3.5cm]{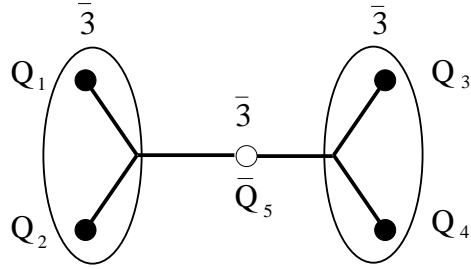} 
\caption{
A QQ-$\rm \bar Q$-QQ type pentaquark configuration.
In the 5Q system, 
$\rm (Q_1, Q_2)$ and $\rm (Q_3, Q_4)$ 
form $\bar {\bf 3}$ representation of SU(3) color, respectively.
The lattice QCD results indicate the multi-Y-shaped flux-tube formation 
in the QQ-$\rm \bar Q$-QQ system.
}
\end{center}
\vspace{-0.2cm}
\end{figure}

We show in Fig.9 the lattice QCD results of the 5Q potential $V_{\rm 5Q}$ 
for symmetric planar 5Q configurations as shown in Fig.8, 
where each 5Q system is labeled by 
$d\equiv \overline{{\rm Q}_1{\rm Q}_2}/2$ and $h\equiv 
\overline{{\rm Q}_1{\rm Q}_3}$.

\begin{figure}[h]
\centering
\hspace{-0.2cm}
\includegraphics[height=3.6cm]{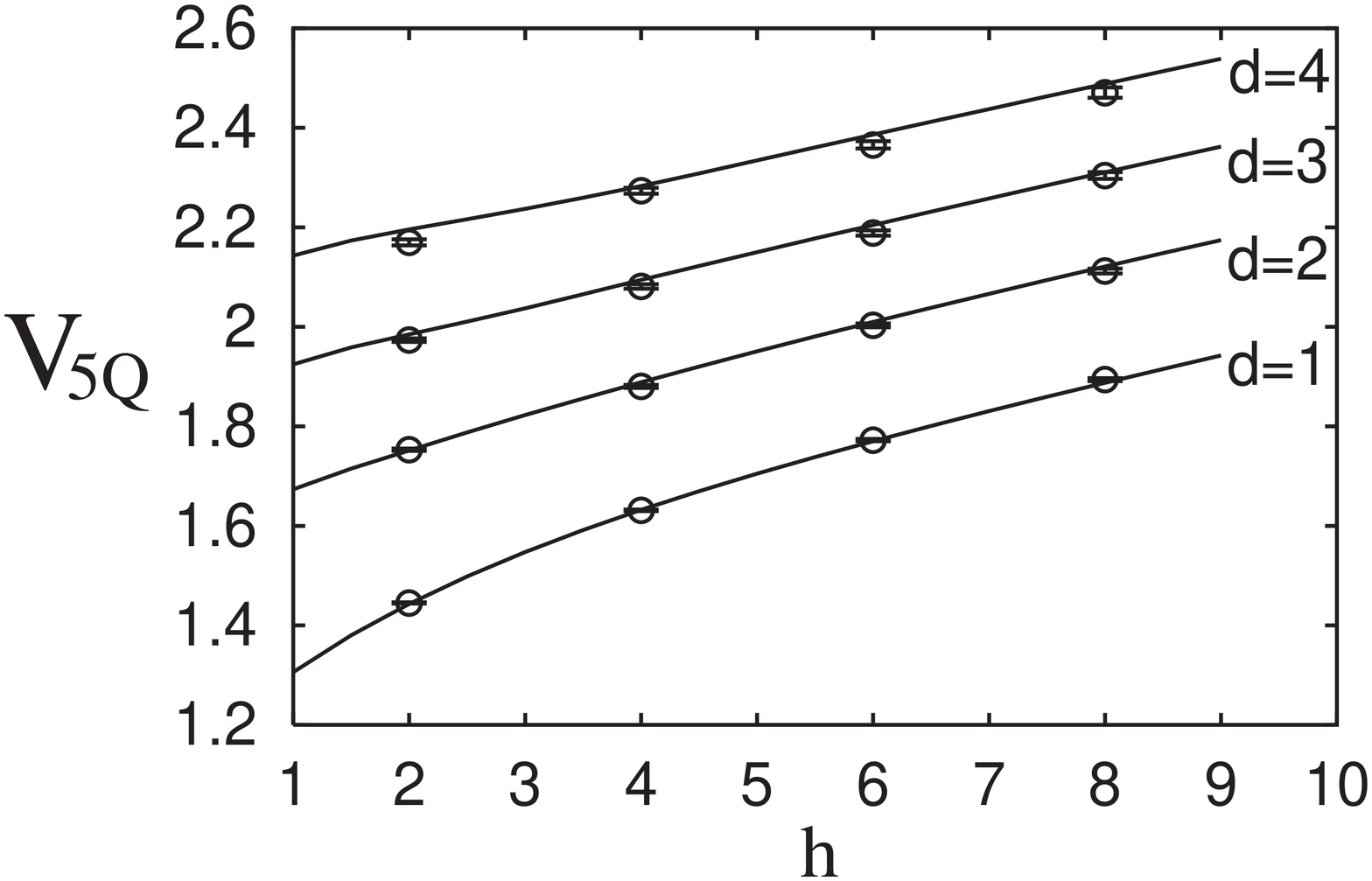}
\includegraphics[height=3.6cm]{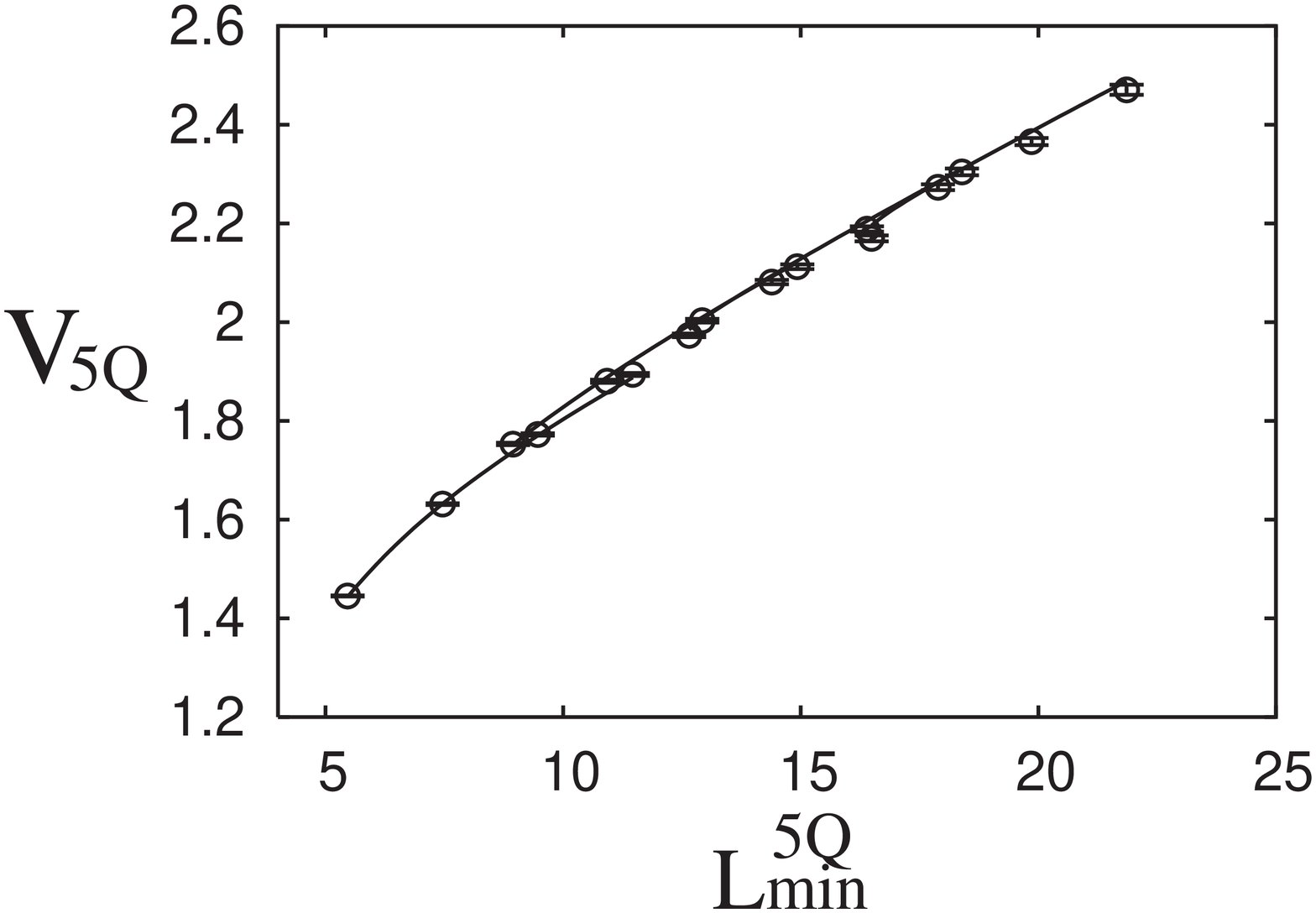}
\caption{Lattice QCD results of the pentaquark potential $V_{\rm 5Q}$ 
for symmetric planar 5Q configurations in the lattice unit:  
(a) $V_{\rm 5Q}$ v.s. $(d,h)$ and (b) $V_{\rm 5Q}$ v.s. $L_{\rm min}^{\rm 5Q}$.
The symbols denote the lattice QCD data, and the curves the theoretical form 
of the OGE plus multi-Y Ansatz.
}
\end{figure}

\noindent
In Fig.9, we add the theoretical curves of  
the OGE Coulomb plus multi-Y Ansatz, where the coefficients 
$(A_{\rm 5Q},\sigma_{\rm 5Q})$ are set to be 
$(A_{\rm 3Q},\sigma_{\rm 3Q})$ obtained from the 3Q potential.\cite{TSNM02}
(Note that there is no adjustable parameter in the theoretical Ansatz  
apart from an irrelevant constant.)
In Fig.9, one finds a good agreement between 
the lattice QCD data of $V_{\rm 5Q}$ and the theoretical curves of 
the OGE Coulomb plus multi-Y Ansatz.

In this way, the pentaquark potential $V_{\rm 5Q}$ is found to be well described 
by the OGE Coulomb plus multi-Y Ansatz as~\cite{OST05,STOI05,OST05p,SOTI05} 
\beq
V_{\rm 5Q}=&-&A_{\rm 5Q}\{ ( \frac1{r_{12}}  + \frac1{r_{34}}) 
+\frac12(\frac1{r_{15}} +\frac1{r_{25}} +\frac1{r_{35}} +\frac1{r_{45}}) \nonumber \\
&+&\frac14(\frac1{r_{13}} +\frac1{r_{14}} +\frac1{r_{23}} +\frac1{r_{24}}) \}
+\sigma_{\rm 5Q}L_{\rm min}^{\rm 5Q}+C_{\rm 5Q}, 
\label{V5Q}
\end{eqnarray}
where $r_{ij}$ 
is the distance between ${\rm Q}_i$ and ${\rm Q}_j$. 
Here, $L_{\rm min}^{\rm 5Q}$ is the minimal total length of the 
flux tube, which is multi-Y-shaped as shown in Fig.8.
This lattice result supports the flux-tube picture for the 5Q system.

\subsection{Tetraquark Potential and Flip-Flop in Lattice QCD}

We study the tetraquark potential $V_{\rm 4Q}$ in lattice QCD 
for about 200 different patterns of QQ-${\rm \bar{Q}\bar{Q}}$ configurations, 
and find the following results.\cite{STOI05,OST05p,OST04}
\begin{enumerate}
\item
When QQ and $\rm \bar Q \bar Q$ are well separated,  
the 4Q potential $V_{\rm 4Q}$ is well described by the OGE Coulomb plus multi-Y Ansatz,
which indicates the multi-Y-shaped flux-tube formation as shown in Fig.10(a).
\item
When the nearest quark and antiquark pair is spatially close, 
the 4Q potential $V_{\rm 4Q}$ is well described by the sum of two Q$\bar {\rm Q}$ potentials, 
which indicates a ``two-meson" state as shown in Fig.10(b).
\end{enumerate}

\begin{figure}[h]
\vspace{-0.5cm}
\begin{center}
\includegraphics[height=3.5cm]{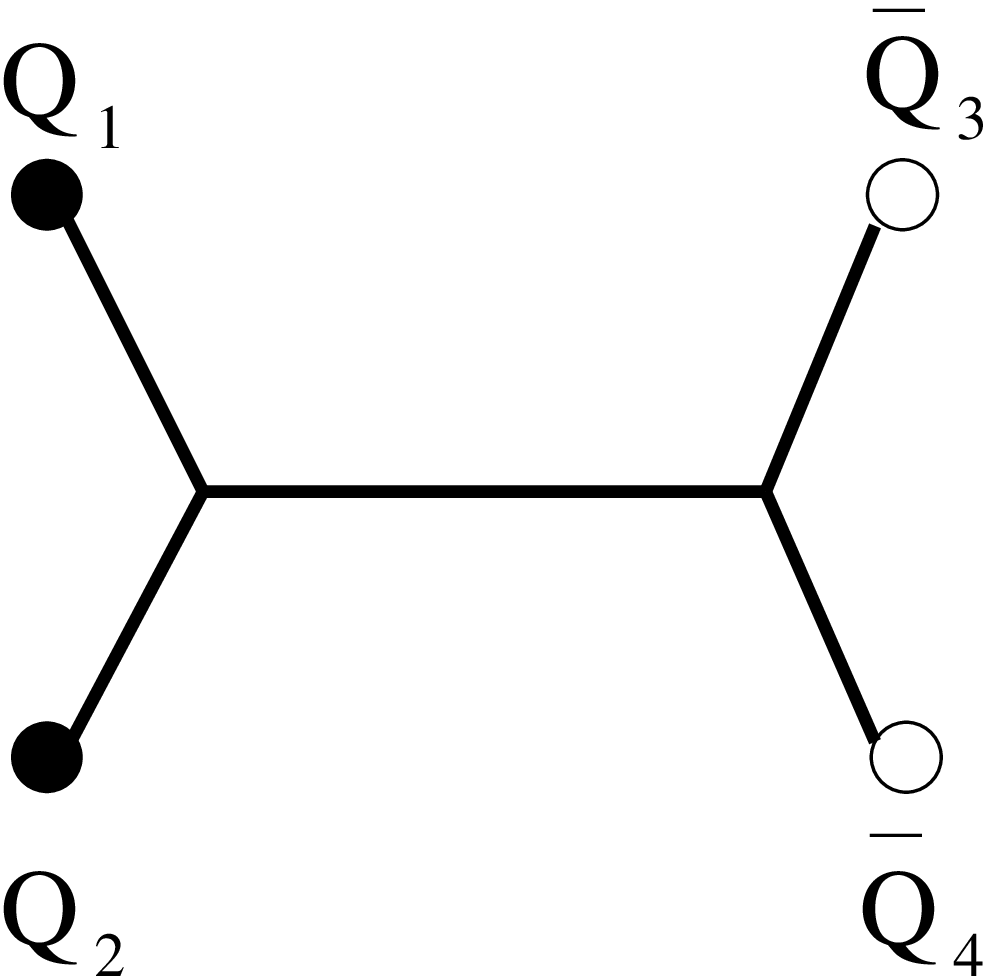}
\hspace{1.5cm}
\includegraphics[height=3.2cm]{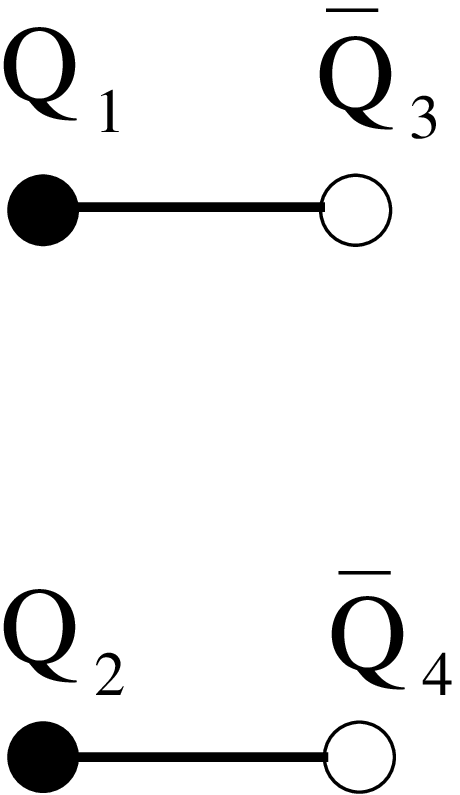} 
\caption{
(a) A connected tetraquark (QQ-$\rm \bar Q\bar Q$) configuration 
and (b) A ``two-meson" configuration.
The lattice QCD results indicate the multi-Y-shaped flux-tube formation 
for the connected 4Q system.
}
\end{center}
\vspace{-0.4cm}
\end{figure}

We show in Fig.11 the lattice QCD results of the 4Q potential $V_{\rm 4Q}$ 
for symmetric planar 4Q configurations as shown in Fig.10, 
where each 4Q system is labeled by 
$d\equiv \overline{{\rm Q}_1{\rm Q}_2}/2$ and $h\equiv 
\overline{{\rm Q}_1{\rm Q}_3}$.

\begin{figure}[h]
\vspace{-0.3cm}
\centering
\includegraphics[height=3.8cm,clip]{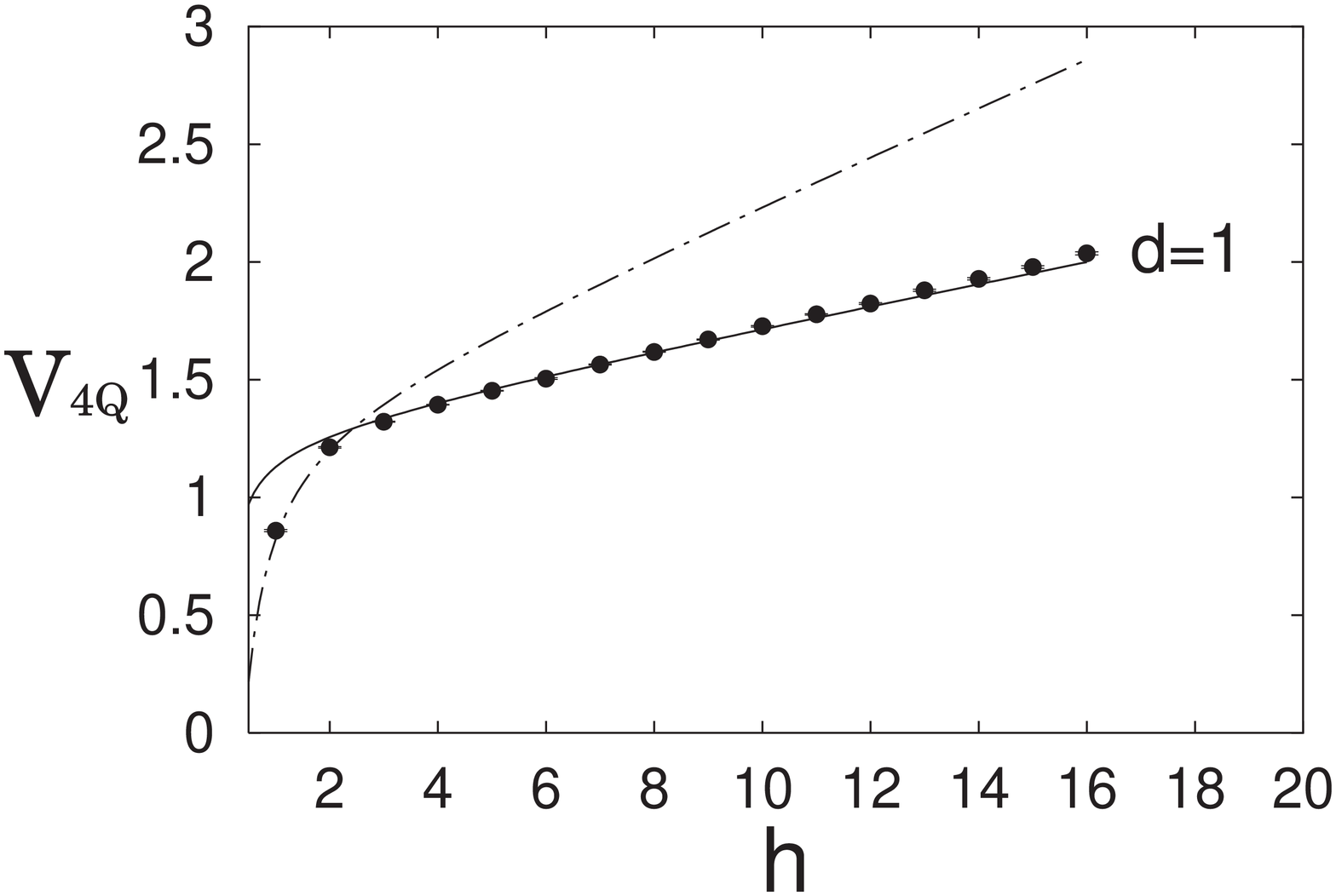}
\hspace{0.1cm}
\includegraphics[height=3.8cm,clip]{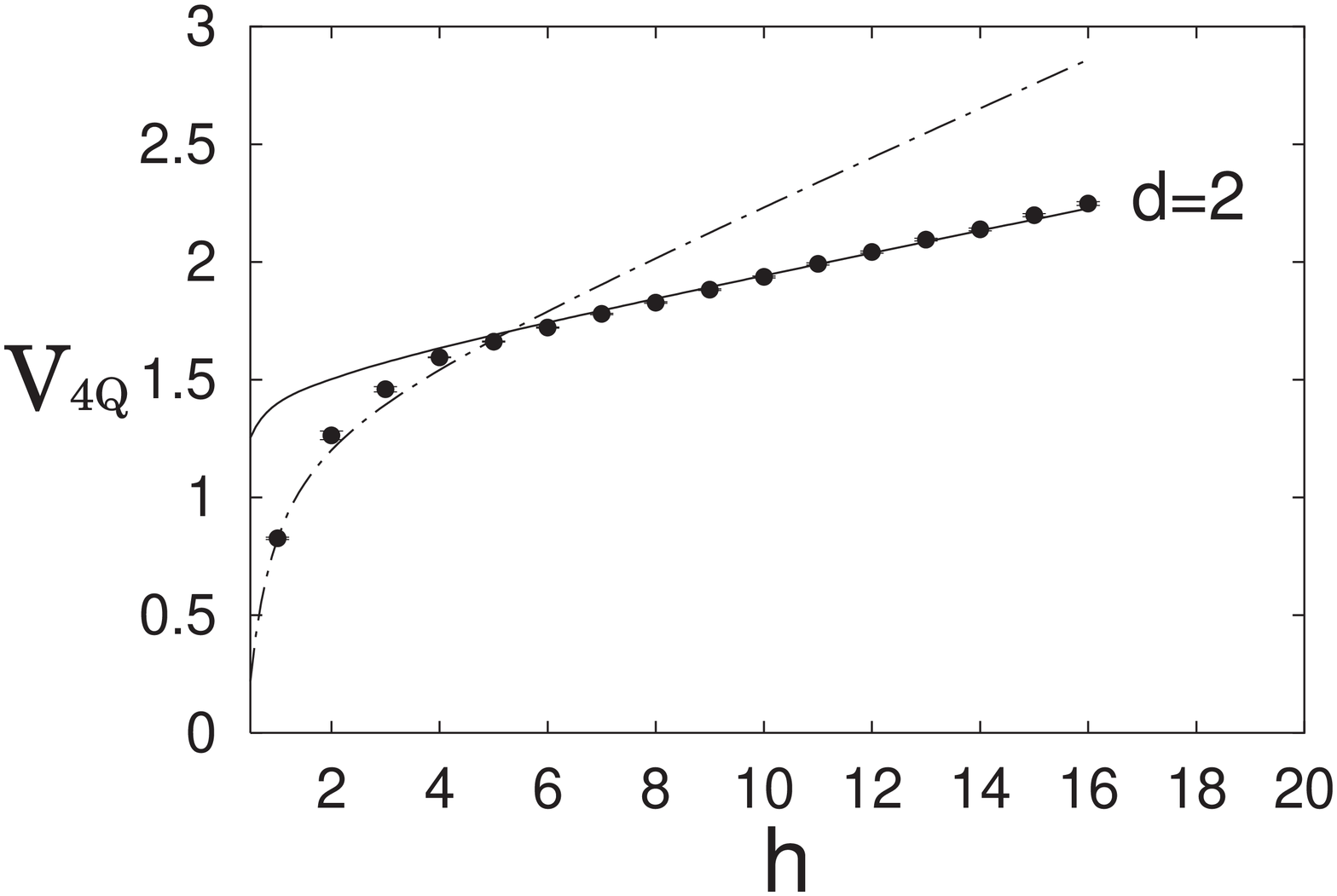}
\caption{
Lattice QCD results of the tetraquark potential $V_{\rm 4Q}$ 
for symmetric planar 4Q configurations in the lattice unit.
The symbols denote the lattice QCD data. 
The solid curve denotes the OGE plus multi-Y Ansatz, 
and the dotted-dashed curve the two-meson Ansatz.}
\end{figure}

For large value of $h$ compared with $d$, the lattice data seem to coincide with 
the solid curve of the OGE Coulomb plus multi-Y Ansatz,
\beq
V_{\rm 4Q}
&=&-A_{\rm 4Q}\left\{ \left( \frac{1}{r_{12}}+\frac{1}{r_{34}} \right)+\frac1{2}\left( \frac{1}{r_{13}}+\frac{1}{r_{14}}+\frac{1}{r_{23}}+\frac{1}{r_{24}} \right) \right \} \nonumber \\
&&+\sigma_{\rm 4Q}L_{\rm min}^{\rm 4Q}+C_{\rm 4Q},
\label{V4Q}
\eeq
where $L_{\rm min}^{\rm 4Q}$ is the minimal total length of 
the flux tube, which is multi-Y-shaped as shown in Fig.10(a). 
Here, the coefficients $(A_{\rm 4Q},\sigma_{\rm 4Q})$ are set to be 
$(A_{\rm 3Q},\sigma_{\rm 3Q})$ obtained from the 3Q potential.\cite{TSNM02}

For small $h$, 
the lattice data tend to agree with 
the dotted-dashed curve of the ``two-meson" Ansatz, 
where the 4Q potential is described 
by the sum of two Q$\bar {\rm Q}$ potentials as 
$V_{\rm Q\bar Q}(r_{13})+V_{\rm Q\bar Q}(r_{24})=2V_{\rm Q\bar Q}(h)$. 

Thus, the tetraquark potential $V_{\rm 4Q}$ is found to take 
the smaller energy of the connected 4Q state or the two-meson state.
In other words, we observe a clear lattice QCD evidence 
of the ``flip-flop", i.e., the flux-tube recombination 
between the connected 4Q state and the two-meson state.
This lattice result also supports the flux-tube picture for the 4Q system.

\subsection{Proper Quark-Model Hamiltonian for Multi-Quarks}

From a series of our lattice QCD studies\cite{TMNS01,TSNM02,TS0304,OST05,STOI05,OST05p,SOTI05,OST04} 
on the inter-quark potentials, 
the inter-quark potential is clarified to consist of 
the one-gluon-exchange (OGE) Coulomb part  
and the flux-tube-type linear confinement part 
in Q$\rm \bar Q$-mesons, 3Q-baryons and multi-quark (4Q, 5Q) hadrons. 

Furthermore, from the comparison among 
the $\rm Q\bar Q$, 3Q, 4Q and 5Q potentials in lattice QCD,
we find the universality of the string tension $\sigma$,
\beq
\sigma_{\rm Q\bar{\rm Q}}\simeq \sigma_{\rm 3Q} \simeq \sigma_{\rm 4Q} 
\simeq \sigma_{\rm 5Q},
\eeq
and the OGE result of the Coulomb coefficient $A$ as 
\beq
\frac1{2}A_{\rm Q\bar{\rm Q}}\simeq A_{\rm 3Q}
\simeq A_{\rm 4Q}\simeq A_{\rm 5Q}
\eeq
in Eqs.(\ref{VQQ}), (\ref{V3Q}), (\ref{V5Q}) and (\ref{V4Q}).

Here, the OGE Coulomb term is considered to originate 
from the OGE process, which plays the dominant role at short distances, 
where perturbative QCD is applicable.
The flux-tube-type linear confinement would 
be physically interpreted by the flux-tube picture, where 
quarks and antiquarks are linked by the one-dimensional squeezed color-electric flux tube  
with the string tension $\sigma$. 

To conclude, the inter-quark interaction would 
be generally described by the sum of the short-distance two-body OGE part  
and the long-distance flux-tube-type linear confinement part 
with the universal string tension 
$\sigma \simeq 0.89$ GeV/fm.

Thus, based on the lattice QCD results, we propose 
the proper quark-model Hamiltonian $\hat H$ for multi-quark hadrons as 
\beq
\hat H=\sum_{i} \sqrt{\hat {\bf p}_i^2+M_i^2}+\sum_{i<j}V_{\rm OGE}^{ij}
+\sigma L_{\rm min},
\eeq
where $L_{\rm min}$ is the minimal total length of the flux tube linking quarks. 
$V_{\rm OGE}^{ij}$ denotes the OGE potential between $i$th and $j$th quarks,
which becomes the OGE Coulomb potential in Eq.(\ref{VnQ}) in the static case.
$M_i$ denotes the constituent quark mass.
The semi-relativistic treatment would be necessary for light quark systems.

It is desired to investigate various properties of multi-quark hadrons 
with this QCD-based quark model Hamiltonian $\hat H$.

\section{Summary and Concluding Remarks}

We have studied tetraquark and pentaquark systems in lattice QCD, 
motivated by the recent experimental discoveries of multi-quark candidates such as 
the $\Theta^+(1540)$.

First, we have performed accurate mass calculations 
of low-lying 5Q states with $J=1/2$ and $I=0$ 
in both positive- and negative-parity channels in anisotropic lattice QCD. 
We have found that the lowest positive-parity 5Q state has 
a large mass of about 2.24GeV after the chiral extrapolation. 
To single out the compact 5Q state from $NK$ scattering states, 
we have developed a new method with the hybrid-boundary condition (HBC), 
and have found no evidence of the compact 5Q state below 1.75GeV 
in the negative-parity channel. 

Second, we have performed the first study of the multi-quark potential in lattice QCD 
to clarify the inter-quark interaction in multi-quark systems. 
We have found that 
the 5Q potential $V_{\rm 5Q}$ for the QQ-${\rm \bar{Q}}$-QQ system 
is well described by the ``OGE Coulomb plus multi-Y Ansatz": 
the sum of the one-gluon-exchange (OGE) Coulomb term 
and the multi-Y-type linear term based on the flux-tube picture. 
The 4Q potential $V_{\rm 4Q}$ for the QQ-${\rm \bar{Q}\bar{Q}}$ system 
is also described by the OGE Coulomb plus multi-Y Ansatz, 
when QQ and $\rm \bar Q \bar Q$ are well separated.  
On the other hand, 
the 4Q system is described as a ``two-meson" state with disconnected flux tubes, 
when the nearest quark and antiquark pair is spatially close.
We have observed a lattice-QCD evidence for the ``flip-flop'', i.e., 
the flux-tube recombination between the connected 4Q state and the ``two-meson'' state.
On the confinement mechanism, 
we have clarified the flux-tube-type linear confinement in multi-quark hadrons.
Finally, we have proposed the proper quark-model Hamiltonian 
based on the lattice QCD results.

\vspace{0.5cm}

\noindent
{\bf \large Acknowledgements}: 
F.O. thanks the organizers of QNP05 for their warm hospitality in Korea.
The lattice QCD Monte Carlo calculations have been performed on NEC-SX5 at Osaka University and SR8000 at KEK.

\end{document}